\documentclass[twocolumn,amsmath,amssymb,prb,longbibliography]{revtex4-2}
\usepackage{graphicx}% Include figure files
\usepackage{dcolumn}% Align table columns on decimal point
\usepackage{bm}% bold math
\usepackage{color}  

\begin{document} 

%\bibliographystyle{apsrev}
%\bibliographystyle{naturemag}

%\preprint{}

%\title{\ST{Multipolarized directional emission of silicon vacancy spin qubits in 6H-SiC \\ 
%Multipolarized emission due to inverted excited states of spin qubits in 6H-SiC}} 

\title{Inverted fine structure of a 6H-SiC qubit enabling robust spin-photon interface}
%\title{Inverted fine structure of a 6H-SiC qubit \ST{enabling multipolarized emisson with} robust spin-photon interface}

\author{I.~D.~Breev$^{1,5}$}
\author{Z.~Shang$^{2,3,5}$}
\author{A.~V.~Poshakinskiy$^{1}$}
\author{H.~Singh$^{4}$}
\author{Y.~Berenc{\'e}n$^{2}$}
\author{M.~Hollenbach$^{2,3}$}
\author{S.~S.~Nagalyuk$^{1}$}
\author{E.~N.~Mokhov$^{1}$}
\author{R.~A. Babunts$^{1}$}
\author{P.~G.~Baranov$^{1}$}
\author{D.~Suter$^{4}$}
\author{S.~A.~Tarasenko$^{1}$}
\author{G.~V.~Astakhov$^{2}$}
\email[E-mail:~]{g.astakhov@hzdr.de} 
\author{A.~N.~Anisimov$^{1}$}
\email[E-mail:~]{aan0100@gmail.com}

\affiliation{$^1$Ioffe Institute, Polytechnicheskaya 26, 194021 St.~Petersburg, Russia  \\ 
$^2$Helmholtz-Zentrum Dresden-Rossendorf, Institute of Ion Beam Physics and Materials Research, Bautzner Landstr. 400, 01328 Dresden, Germany  \\
$^3$Technische Universit\"{a}t Dresden, 01062 Dresden, Germany \\
$^4$Fakult\"at Physik, Technische Universit\"at Dortmund, 44221 Dortmund, Germany \\ 
$^5$These authors contributed equally to this work}

\begin{abstract}  
Optically controllable solid-state spin qubits are one of the basic building blocks for applied quantum technology. Efficient extraction of emitted photons and a robust spin-photon interface are crucial for the realization of quantum sensing protocols and essential for the implementation of quantum repeaters. Though silicon carbide (SiC) is a very promising material platform hosting  highly-coherent silicon vacancy spin qubits, a drawback for their practical application is the unfavorable ordering of the electronic levels in the optically excited state. Here, we demonstrate that due to polytypism of SiC, a particular type of silicon vacancy qubits in 6H-SiC possesses an unusual inverted fine structure. This results in the directional emission of light along the hexagonal crystallographic axis, making photon extraction more efficient and integration into photonic structures technologically straightforward. From the angular polarization dependencies of the emission, we reconstruct the spatial symmetry and determine the optical selection rules depending on the local deformation and spin-orbit interaction, enabling direct implementation of robust spin-photon entanglement schemes. Furthermore, the inverted fine structure leads to unexpected behavior of the spin readout contrast. It vanishes and recovers with lattice cooling due to two competing optical spin pumping mechanisms. Our experimental and theoretical approaches provide a deep insight into the optical and spin properties of atomic-scale qubits in SiC required for quantum communication and distributed quantum information processing. 
\end{abstract}

\date{\today}

\maketitle
%---------------------------------------------------------------

%\section{Introduction}
 
Optically-interfaced solid-state spins are considered as candidates for the realization of quantum networks and photonic quantum computing \cite{Atature:2018hh, Awschalom:2018ic}. The practical realization of quantum repeaters requires a system with (i) a high-fidelity spin photon interface, (ii) a source of spectrally indistinguishable single photons and (iii) a long-lived quantum memory. One of the promising candidates are III-V semiconductor quantum dots (QDs) \cite{DeGreve2016}. They are the brightest solid-state source of single photons in the telecom wavelength, which can be used for quantum key distribution. Due to the optical selection rules, there are robust, high-fidelity protocols for the entanglement generation between the photon polarization and the spin state. Yet, a short spin coherence time and a large inhomogeneous broadening of the emission wavelength from individual QDs are the main obstacles for their practical use. Another promising material platform is based on color centers in diamond. Indeed, high-fidelity (92\%) entanglement at a distance of 1.3~km has been demonstrated with two nitrogen-vacancy (NV) centers in diamond \cite{Hensen2015}. The main obstacle for scaling up this system is its spectrally unstable emission, resulting in a very low entanglement generation rate. Another color center in diamond, the silicon-vacancy (SiV), is spectrally stable allowing memory-enhanced quantum communication protocols at mK temperatures \cite{Bhaskar:2020gh}. 

There is a continuous search for other solid-state platforms for integrated quantum photonics with spin qubits. Silicon carbide (SiC) holds great promise as technologically mature material \cite{Castelletto:2020kr, Lukin:2020fa, Son:2020kh}. Particularly, the negatively charged silicon vacancy ($\mathrm{V_{Si}}$) in SiC reveals appealing quantum properties \cite{Baranov:2011ib, Riedel:2012jq}. They can be used as quantum emitters  \cite{Kraus:2013di, Widmann:2014ve, Fuchs:2015ii}  and have an extremely long spin coherence time  \cite{Widmann:2014ve, Simin:2017iw, Soltamov:2019hr}. The $\mathrm{V_{Si}}$ centers can be naturally integrated into photonic structures \cite{Radulaski:2017ic, Bracher:2017il, Lukin:2019fh} due to the established spin-photon interface  \cite{Nagy:2019fw, Udvarhelyi:2019eh} and high spectral stability of their zero-phonon lines (ZPLs) \cite{Banks:2019je, Morioka:2020iv} along with nanoscale engineering of single $\mathrm{V_{Si}}$ with focused ion beams  \cite{Kraus:2017cka, Wang:2017fb}  and the ability for the Stark tuning of the $\mathrm{V_{Si}}$ ZPL \cite{Ruhl:2019hs, Lukin:2020jb}. These achievements pave the way for on-demand generation of indistinguishable single-photon emitters. Furthermore, spin-photon entanglement schemes based on the excited state (ES) fine structure and spin-dependent optical transitions to the ground state (GS) have been theoretically proposed \cite{Soykal:2015uw, Economou:2016bp} but not yet realized. 

The most studied polytype is 4H-SiC, which is characterized by two non-equivalent lattice sites for Si. Correspondingly, there are two $\mathrm{V_{Si}}$ centers in 4H-SiC which are historically labeled as V1 and V2. They differ from each other by their ZPL spectral position and zero-field splitting parameter $2D$ \cite{Wagner:2000fj, Baranov:2011ib}. It has also been established that the V1 ZPL in the low-temperature photoluminescence (PL) spectra splits in two lines V1 and V1', which are orthogonally polarized \cite{Wagner:2000fj, Janzen:2009ij}. The dichroic properties of the V1/V1' ZPL as well as the V2  ZPL in 4H-SiC have been recently investigated in detail  \cite{Nagy:2018ey}. Both the V1 ZPL, which dominates at low temperatures, and the V2 ZPL are linearly polarized along the $c$-axis of the 4H-SiC crystal, indicating that these centers emit light perpendicularly to the $c$-axis. Because the quantum-grade SiC epitaxial layers are usually grown along the $c$-axis, this makes light collection from the growth surface and, consequently, photon extraction from planar photonic structures less efficient. On the other hand, the fabrication of photonic structures from the side of the epitaxial layer is technologically laborious.  

\begin{figure}[t]
\includegraphics[width=.48\textwidth]{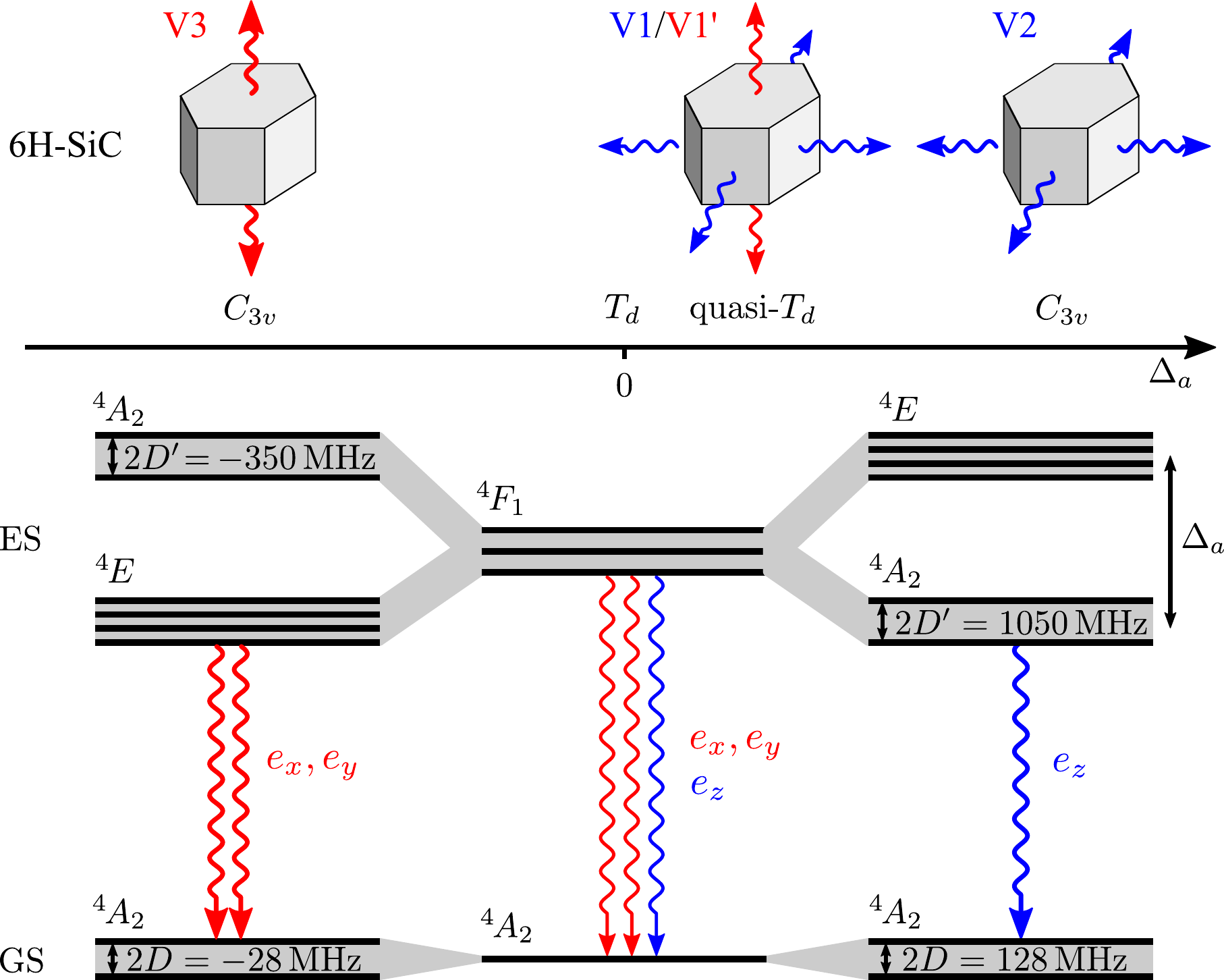}
\caption{Scheme of the $\mathrm{V_{Si}}$ ground state (GS) and excited state (ES) multiplets and polarized directional
emission linked to the ES structure in 6H-SiC. The zero-field splittings of the GS and ES $^4 A_2$ octuplets for the V2 and V3 centers  are obtained from the experiment as described in the text. } \label{fig1} 
\end{figure}

The polytype 6H-SiC has three non-equivalent lattice sites for Si, resulting in three $\mathrm{V_{Si}}$ centers. 
In the present paper, we perform comprehensive study of these $\mathrm{V_{Si}}$ centers by polarization-resolved optical spectroscopy and optically-detected magnetic resonance (ODMR) spectroscopy.
The V1/V1' and V2 centers in 6H-SiC have very similar properties those in 4H-SiC. 
In contrast, the V3 center in 6H-SiC is optically active in two perpendicular polarizations and emits light preferentially along the $c$-axis, making the photon extraction from planar photonic structures easier. We explain this property by the inverted fine structure of the excited state with respect to other $\mathrm{V_{Si}}$ centres in 4H- and 6H-SiC. It also enables an entanglement scheme between the circular polarization of the emitted photon and the $\mathrm{V_{Si}}$ spin 
\cite{Economou:2016bp}, 
which is not possible for the V1 or V2 centers. 
We also find that the V3 center has a unusual temperature-induced 
inversion of the ODMR signal at a critical point $T_c = 16 \, \mathrm{K}$.    

The V1/V1', V2, and V3 centers differ by their deviation from the cubic symmetry $T_d$ to the trigonal symmetry $C_{3v}$. This deviation can be characterized by the axial splitting $\Delta_a$ between the $^4 A_2$ quadruplet and $^4 E$ octuplet of the spin-3/2 ES, as depicted in Fig.~\ref{fig1}. For cubic symmetry, $\Delta_a =0$ and the excited state is then the 12-fold $^4 F_1$ multiplet.
For $\Delta_a  \neq 0$,  the $^4 F_1$ multiplet splits into the $^4 A_2$ quadruplet and $^4 E$ octuplet, with the order being determined
by the sign of $\Delta_a$. Depending on what multiplet has the lowest energy, the PL of the center is polarized either along ($e_z$) or perpendicular ($e_x$, $e_y$) to the $c$-axis, as indicated by the wavy arrows in Fig.~\ref{fig1}.  
As compared to V1 and V2, the V3 center has an inverted ES structure, which determines its unusual multipolarized ($e_x$, $e_y$) optical emission.

\section{Inverted excited-state structure} 

\subsection{PL polarization}   

\begin{figure*}[t]
\includegraphics[width=.8\textwidth]{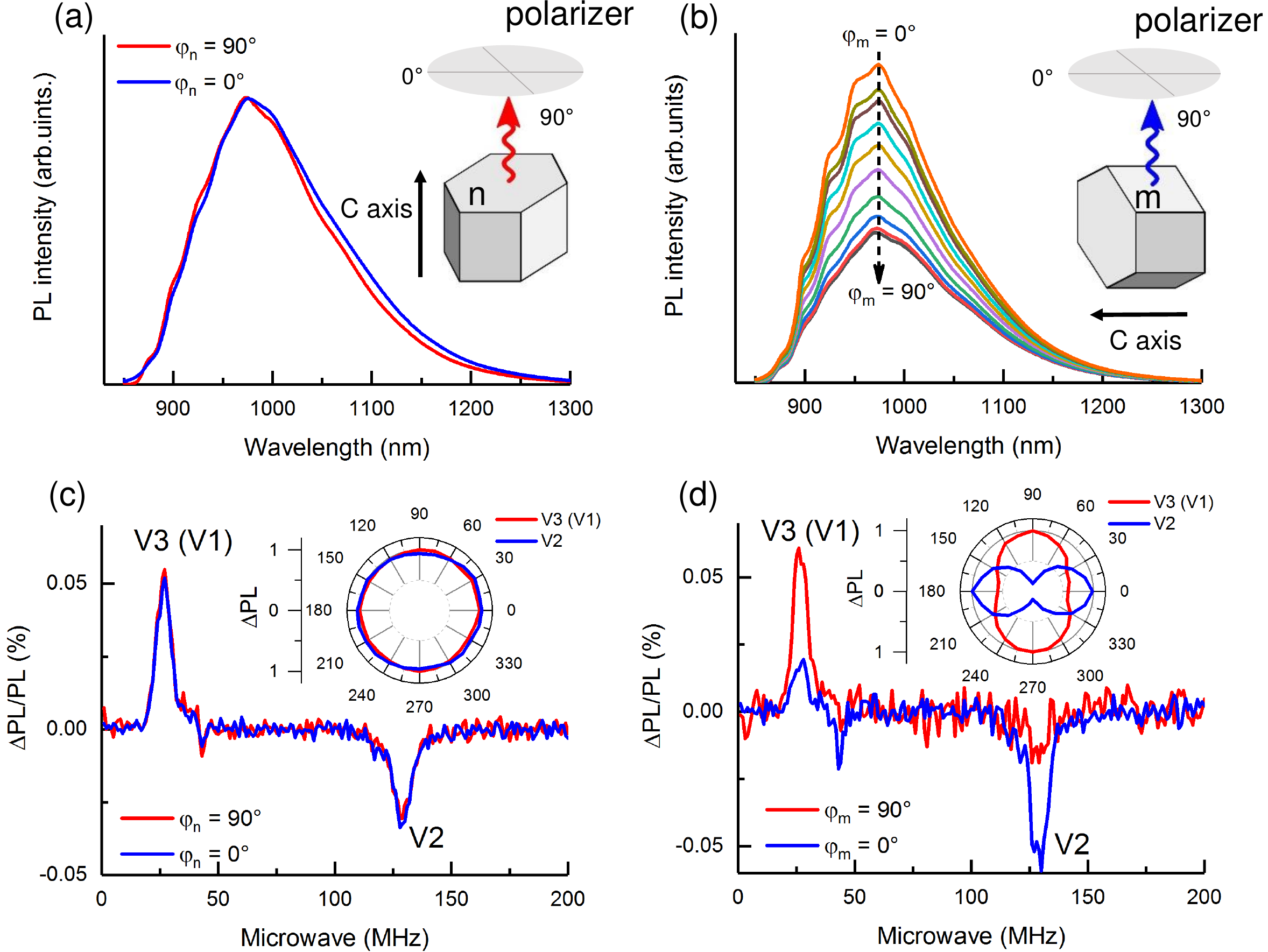}
\caption{Room-temperature polarization dependence  of the $\mathrm{V_{Si}}$  PL in 6H-SiC. (a), (b) 
PL spectra of $\mathrm{V_{Si}}$ centers taken from the $n$-face (perpendicular to the $c$-axis) and $m$-face (parallel to the $c$-axis), respectively, collected through the linear polarizer. The insets sketch the experimental geometries and the polarizer orientations. (c), (d)  ODMR spectra of the V3 (V1) and V2 $\mathrm{V_{Si}}$  centers at $\varphi_{n,m} = 0^{\circ}$ (blue curve) and $\varphi_{n,m} = 90^{\circ}$ (red curve) of the polarizer orientation. Insets: normalized polar plots $\mathrm{\Delta PL}(\varphi_{n,m})$ for the $n$-face (c) and $m$-face (d), respectively. } \label{fig2}
\end{figure*}
 
First, we measure the orientation dependence of the polarization ${V_\mathrm{Si}}$ PL in 6H-SiC at room temperature (Fig.~\ref{fig2}). The PL collected from the $n$-face, i.e., along the $c$-axis, is unpolarized (Fig.~\ref{fig2}(a)). In contrast, the PL collected from the $m$-face, i.e., perpendicular to the $c$-axis, is partially polarized along the $c$-axis (Fig.~\ref{fig2}(b)). At room temperature, the emission from the different ${V_\mathrm{Si}}$ centers overlap spectrally and it is not possible to distinguish between them. To separate spectral contributions from the different ${V_\mathrm{Si}}$ centers, we the microwave (MW) assisted spectroscopy \cite{Shang:2021bq}.  Figure~\ref{fig2}(c) and (d) present ODMR spectra with a peak at $28 \, \mathrm{MHz}$ corresponding to the V3 (probably also V1) ${V_\mathrm{Si}}$ centers and a negative peak at $128 \, \mathrm{MHz}$ corresponding to the V2 ${V_\mathrm{Si}}$ center \cite{Baranov:2011ib}.  The V2 $\mathrm{\Delta PL / PL}$ is clearly linearly polarized as 
$e_z \| c$ (the inset of  Fig.~\ref{fig2}(d)). The polarization of the V3/V1 centers cannot be separated at room temperature and their cumulative contribution to the $\mathrm{\Delta PL / PL}$ is preferentially polarized along the perpendicular direction as $e_x, e_y  \bot c$. 

\begin{figure*}[t]  
\includegraphics[width=.7\textwidth]{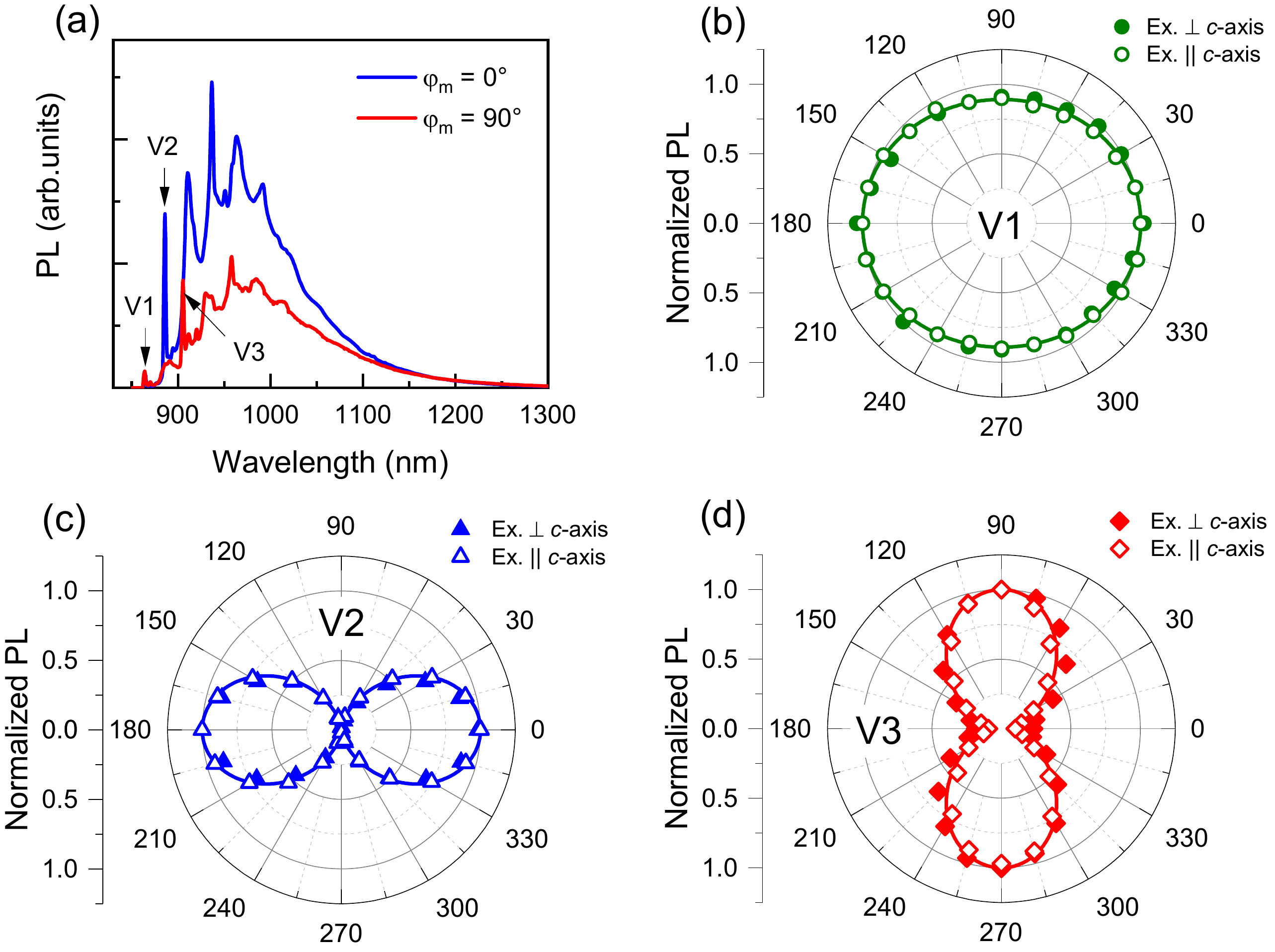}
\caption{Polarization dependencies of the $\mathrm{V_{Si}}$ ZPLs in 6H-SiC at $T = 15 \, \mathrm{K}$. (a) PL spectra measured from the $m$-face at $\varphi_{m} = 0^{\circ}$ (blue curve) and $\varphi_{m} = 90^{\circ}$ (red curve) of the polarizer orientation with respect to the $c$-axis.  (b), (c), (d) The polar polarization plots of the V1, V2 and V3 ZPL intensities, respectively. The solid and open symbols are the measured data for the excitation polarization perpendicular and parallel to the $c$-axis. The solid lines are fits to Eq.~(\ref{angular}) with the fitting parameter  $\cos 2\theta$ of $0.06$, $0.96$ and $-0.89$ for the V1, V2 and V3 ZPL, respectively. } \label{fig3}
\end{figure*}

At low temperature ($T = 15 \, \mathrm{K}$), the V1, V2 and V3 ${V_\mathrm{Si}}$  centers in 6H-SiC are spectroscopically distinguishable by their ZPLs \cite{Wagner:2000fj}, as presented in Fig.~\ref{fig3}(a). 
We fit the angular dependencies from the $m$-face in Fig. 3(b)-(d) to
\begin{align}\label{angular}
I_m(\varphi)  =  I_0(1+ \cos 2\theta \,\cos 2\varphi_m), 
\end{align}
where $\varphi_m$ is the angle between the linear polarizer axis and the $c$-axis,  $I_0$ is the average intensity. The phenomenological angle $\theta$ characterizes the components of the matrix element $\bm d$ of the dipole optical transition, $\tan \theta = |d_x|/|d_z| = |d_y|/|d_z|$.  For the pure $^4A_1$ and $^4E$ states, one has $\theta = 0^\circ$ and $90^\circ$, respectively. 

The V1 ZPL shows nearly unpolarized emission (Fig.~\ref{fig3}(b)) and a fit to Eq.~(\ref{angular}) gives $\cos 2\theta = 0.06$ 
($| d_x /d_z | \approx 0.94$). Earlier studies show that it consists of two, V1 and V1' ZPLs, which are polarized as $e_z \| c$ and  $e_x, e_y  \bot c$, respectively, and split by  $1.1 \, \mathrm{meV}$, where V1' is the high-energy state transition \cite{Janzen:2009ij}. In our experiments, we do not resolve the V1 and V1' ZPLs and therefore, the cumulative PL emission associated with the V1 ${V_\mathrm{Si}}$ center is nearly unpolarized.  

After the background subtraction (Supplementary Information), the V2 ZPL shows nearly 100\% polarization with $e_z \| c$ (Fig.~\ref{fig3}(c)).  A fit to Eq.~(\ref{angular}) gives $\cos 2\theta = 0.96$ ($\theta=  8^\circ \approx 0^\circ$). The V3 ZPL is orthogonally polarized $e_x, e_y  \bot c$ with small but not vanishing contribution $e_z \| c$ (Fig.~\ref{fig3}(d)). A fit to Eq.~(\ref{angular}) gives $\cos 2\theta = -0.89$ ($\theta=  76^\circ \approx 90^\circ$). We verify that the observed PL polarization does not depend on the excitation laser polarization (open and solid symbols in Figs.~\ref{fig3}(b-d)). Furthermore, we observe nearly the same angular dependencies at a temperature  $T = 100 \, \mathrm{K}$ (Supplementary Information), indicating that they are not related to the thermal population in the ES, which thus manifest intrinsic properties of the corresponding ${V_\mathrm{Si}}$   centers, as presented in Fig.~\ref{fig1}.

\subsection{Theory}

\begin{figure*}[t] 
\includegraphics[width=.99\textwidth]{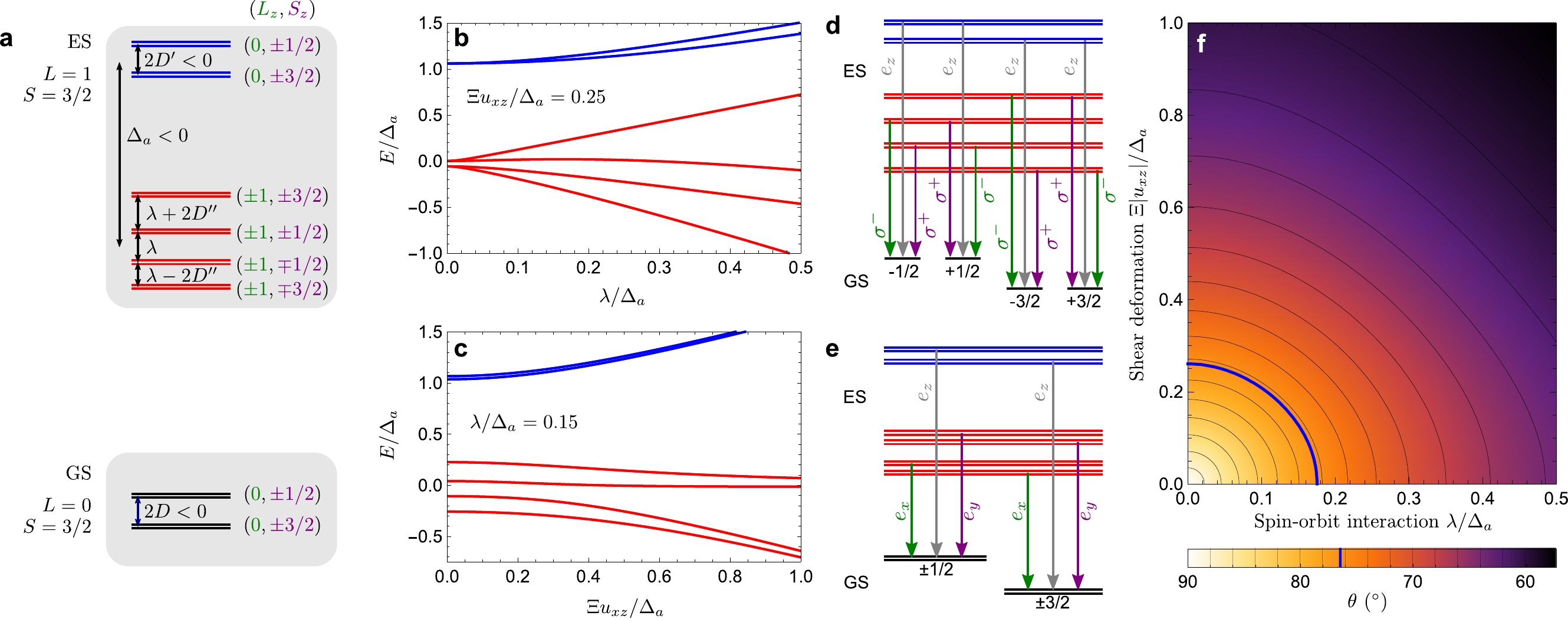}
\caption{Ground state (GS) and excited states (ES) fine structure of the V3 $\mathrm{V_{Si}}$ center in 6H-SiC. (a) Schematic presentation of the GS and ES spin sublevels depending on $L_z$ and $S_z$.  (b) Dependence of the ES spin sublevels on  the spin-orbit interaction strength $\lambda$ in the presence of shear strain $\Xi u_{xz}/\Delta_a = 0.25$. 
(c) Dependence of the ES spin sublevels on the shear strain $u_{xz}$ in the presence of spin-orbit interaction $\lambda/\Delta_a = 0.15$.
(d), (e) Selection rules for the optical transitions between the ES and GS levels shown for the cases of (d) dominant spin-orbit interaction and (e) dominant deformation coupling. 
(f) Color plot of the angle $\theta = \arctan (|d_x/d_z|)$, describing the PL polarization at the transitions from the ${}^4E$ ES multiplet to the GS, as a function of the spin-orbit interaction strength $\lambda$ and the shear strain $u_{xz}$. The thick line indicates the angle $\theta = 76^\circ$ measured for the V3 center in our experiments. }\label{fig4}
\end{figure*}

Here, we construct the effective ES spin Hamiltonian, calculate the optical selection rules and compare them with 
the polarization dependencies observed in the experiments. 
The spatial symmetry of the vacancy-related spin centers in hexagonal SiC is described by the $C_{3v}$ point group~\cite{Simin:2016cp}. This group has 
one-dimensional irreducible representations $A_1$ and $A_2$ and a two-dimensional irreducible representation $E$. Therefore, the all $\mathrm{V_{Si}}$  spin-3/2 spectral terms are either 4-fold or 8-fold multiplets. It is natural to assume that the $V_{\rm Si}$ GS in 6H-SiC is orbitally nondegenerate and corresponds to the $^4A_2$ quadruplet, similarly to the $V_{\rm Si}$ GS in 4H-SiC~\cite{Soykal:2015uw}. 

The low-temperature PL polarization  is determined by the lowest ES multiplet which can be either $^4A_2$ quadruplet or $^4E$ octuplet~\cite{Soykal:2015uw}.
The splitting between the $^4A_2$ and $^4E$ multiplets is caused by the distortion of the $V_{\rm Si}$ symmetry from the cubic $T_d$ symmetry. 
In the $T_d$ approximation, this splitting vanishes and the $^4A_2$ quadruplet and the $^4E$ octuplet merge into the 12-fold 
$^4 F_1$ multiplet of the $T_d$ group, as shown in  Fig.~\ref{fig1}.

The ES-GS transitions $^4A_2 \rightarrow ^4A_2$ are optically active in the polarization $e_z$ ($z \parallel c$). The PL collected from the hexagonal 
$n$-plane should be quite weak,while the intensity of PL emitted from the side $m$-plane and transmitted through the polarizer should vary as
 $I \propto \cos^2 \varphi_m$. 
 Such a polarization dependence is observed for the V2 ZPL (Figs.~\ref{fig2} and~\ref{fig3}) and indicates that the lowest excited multiplet of 
 the V2 center is $^4A_2$. 
 
Contrary, the ES-GS  transitions $^4E \rightarrow ^4A_2$ are optically active in the $e_x$ and $e_y$  polarizations, which are perpendicular to the $c$-axis.
The corresponding PL collected from the $n$-plane should be strong and unpolarized (at zero magnetic field) and the PL intensity collected from the $m$-plane through the polarizer should follow the $\varphi$-dependence $I \propto \sin^2 \varphi_m$. This polarization behavior is observed for the V3 ZPL, see Figs.~\ref{fig2} and~\ref{fig3}. Therefore, we conclude that 
the ordering of the ES multiplets is inverted  for the V3 center, with $^4E$ being the lowest multiplet.

The PL at the V1/V1' ZPL is nearly unpolarized indicating that the $^4A_2$ and $^4E$ multiplets are close to each other and that the
effective symmetry of the $V_1$ center is close to $T_d$, as shown in Fig.~\ref{fig1}. 

We note that the measured polarization dependencies at the $V_2$ and $V_3$ ZPLs are not as strict as the orbital symmetry suggests.
This deviation can be attributed to the spin-orbit interaction or to a further distortion of the $V_{\rm Si}$ symmetry from $C_{3v}$ due to local strain or the Jahn-Teller effect and will be discussed later.

To study the ES fine structure and the effect of strain, we construct now the effective Hamiltonian 
of the $^4A_2 \,+\, ^4E$ states.  These 12 states have angular momentum $L=1$ and spin $S=3/2$. 
The effective Hamiltonian can be expressed via the operators of the angular momenta $\bm L$ and $\bm S$ 
\begin{align}\label{H}
H = H_e + H_\text{so} + H_\text{ss} + H_\text{def} \,,
\end{align}
where the four terms correspond to the pure orbital, spin-orbital, spin-spin, and deformation interactions.
The orbital term has the form
\begin{align}
H_e  = \Delta_a \left( L_z^2 - \frac23 \right) 
\end{align}
and describes the splitting $\Delta_a$ between the $^4 E$ octuplet  (states with the eigenvalues $L_z=\pm 1$)
and the $^4 A_2$ quadruplet (states with the eigenvalue $L_z=0$ of ). 
The spin-orbit interaction, linear in the spin operator $\bm S$, has the form
\begin{align}
H_\text{so} = \lambda \, \bm{L} \cdot \bm{S} \,,
\end{align}
where $\lambda$ is the ES spin-orbit interaction constant. 
The spin-spin interaction, quadratic in the spin operators~\cite{LandauLifshitz3}, is given by the sum of two terms
\begin{align}\label{eq:Hss}
H_\text{ss} =  b\left[ (\bm S \cdot \bm L)^2 - \frac52 \right] + \tilde{D} \left(S_z^2 - \frac54 \right) \,,
\end{align}
where $b$ and $\tilde{D}$ are the spin-spin interaction parameters. 
Finally, the deformation mixing of orbital states is described by the Hamiltonian \cite{BirPikus}
\begin{align}\label{eq:Hdef}
H_\text{def} = \Xi_e \sum_{\alpha\beta} \left( u_{\alpha\beta} - \frac{\delta_{\alpha\beta}}{3} \text{Tr\,}u \right)
\left(L_{\alpha}L_{\beta}-\frac23\delta_{\alpha\beta} \right) ,
\end{align}
where $\Xi_e$ is the deformation potential constant, $u_{\alpha\beta}$ is the strain tensor, and  
$\text{Tr\,}u=\sum_\alpha u_{\alpha\alpha}$. The effect of strain on the spin-orbit and spin-spin interactions is much smaller
and neglected here. To avoid multiplication of parameters, all the contributions to the Hamiltonian~\eqref{H} are given 
in the isotropic approximation except for the terms $(L_z^2-2/3)$ and $(S_z^2 - 5/4)$, which vanish in the isotropic model 
and take into account the axial symmetry.

We assume the following energy hierarchy: $\Delta_a \gg \lambda \gg b,\tilde{D}$ and $\Delta_a \gg \Xi_e u_{\alpha\beta}$.
Then, using the L\"{o}wdin perturbation theory \cite{Lowdin}, we obtain the effective Hamiltonian of the $^4 A_2$ quadruplet  
\begin{align}
H_{A_2} = \varepsilon_{A_2} +  D' \left(S_z^2 - \frac54 \right) ,
\end{align} 
where $\varepsilon_{A_2} \approx - (2/3) \Delta_a$ is the quadruplet position and 
$2D'$ is the zero-field splitting between the $\pm1/2$ and $\pm 3/2$ spin sublevels 
of the $^4 A_2$ ES quadruplet. For the first order in the spin-spin interaction and the second order in the spin-orbit interaction,
the zero-field splitting constant has the form 
\begin{align}
D' = \tilde{D} - b + \frac{\lambda^2}{\Delta_a},
\end{align}
for the $^4E$ octuplet the L\"{o}wdin perturbation theory gives the effective Hamiltonian
\begin{align}
&H_{E} = \varepsilon_{E}  + \lambda S_z \sigma_z + D'' \left( S_z^2 - \frac54 \right) \\
&+  D''_\perp \, [ (S_x^2-S_y^2)\sigma_x +  (S_xS_y+S_yS_x)\sigma_y] \nonumber \\
&+ \frac12 \Xi_e [(u_{xx}-u_{yy})\sigma_x+2u_{xy}\sigma_y ]\,, \nonumber
\end{align}
where $\varepsilon_{E} \approx \Delta_a /3$ is the octuplet position,
$\sigma_x$, $\sigma_y$, and $\sigma_z$ are the Pauli matrices in the space of orbital states with $L_z = \pm 1$, 
\begin{align}
&D'' = \tilde{D} + \frac{b}2  -\frac{\lambda^2}{2\Delta_a} \,, \\  
& D''_\perp = \frac{b}2 + \frac{\lambda^2}{2\Delta_a} \,.
\end{align}

Figure~\ref{fig4}(a) shows schematically the GS and ES  multiplet structure in the absence of strain. For $\Delta_a < 0$,  which corresponds to the V3 center, the $^4 E$ octuplet is below the $^4 A_2$ quadruplet in the ES. The spin-orbit and spin-spin interactions form the fine structure of the multiplets. We cannot determine the dominant contribution to the $^4 E$ octuplet splitting from our experimental data. According to DFT calculations, the spin-orbit interaction constant is $\lambda/h \sim 100\,$GHz for the V2 $\mathrm{V_{Si}}$ in 4H-SiC \cite{Udvarhelyi:2020fq}. Another estimate gives $\lambda/h \sim 5\,$GHz \cite{Economou:2016bp}. 
The spin-orbit and deformation interactions not only determine the fine structures of the $^4E$ and $^4A_2$ multiplets but also mix them. Figure~\ref{fig4}(b) shows the ES structure as a function of the spin-orbit interaction strength $\lambda$ in the presence of a small shear strain $u_{xz}$. The energy levels are calculated numerically by the direct diagonalization of Hamiltonian~\eqref{H}. When $\lambda$ becomes comparable to the splitting between the multiplets $\Delta_a$, they get fully mixed. Figure~\ref{fig4}(c) shows the ES structure as a function of the shear strain $u_{xz}$. The deformation coupling of different orbital states suppresses spin-orbit splitting. When $\Xi^2 u_{xz}^2/\Delta_a>>\lambda$, the $^4E$ octuplet transforms into the pair of quadruplets. 

Next, we analyze the polarization of the electric dipole optical transitions between the ES and GS.  In the case of negligible mixing of the ES multiplets, the transitions from the $^4A_2$ quadruplet are active in the $e_z$ polarization while  the transitions from the $^4E$ octuplet are active in the $e_x$ and $e_y$ polarizations. The selection rules for the transitions between particular levels of the $^4E$ ES octuplet and the $^4A_2$ GS quadruplet are determined by the fine structures of the multiplets. If the spin-orbit interaction dominates, $\lambda \gg \Xi u_{xz}$, these transitions are active in the polarizations $(e_x \pm i e_y)/\sqrt{2}$ and occur with the emission of circularly polarized photons, as depicted in Fig.~\ref{fig4}(d). If the deformation interaction induced by shear strain dominates, $\Xi u_{xz} \gg \lambda$, the emitted photons are linearly polarized, Fig.~\ref{fig4}(e).

The strict selection rules, which imply that the optical transitions from the $^4A_2$ and the $^4E$ states are active in the polarizations 
parallel and perpendicular to the $c$-axis, respectively, are violated if the multiplets are mixed. To study this violation, we calculate the phenomenological tilt angle of the electric dipole $\theta = \text{arctan\,} |d_x/d_z|$, which determines the PL polarization, taking into account the mixing of the multiplets by the spin-orbit interaction and strain. Figure~\ref{fig4}(f) shows the angle $\theta$ for the optical transitions from the $^4E$ multiplet as function of the parameters $\lambda / \Delta_a$ and $\Xi u_{xz} / \Delta_a$. The thick solid curve corresponds to the value $\theta_{E} = 76^\circ$ observed in our experiment for the V3 centers (Fig.~\ref{fig3}(d)). In the case of weak mixing of the multiplets, 
$\Xi u_{xz}, \Xi u_{yz}, \lambda \ll \Delta_a$, the dipole angles for the transitions from the $^4A_2$ and the $^4E$ states have the form
\begin{align}
&\theta_{A_2} \approx \frac{\sqrt{\frac{\Xi^2}2( u_{xz}^2 + u_{yz}^2) + \frac54 \lambda^2}}{\Delta_a}  \,, \\
&\theta_{E} \approx \frac{\pi}2 - \frac{\sqrt{\Xi^2( u_{xz}^2 + u_{yz}^2) + \frac52 \lambda^2}}{\Delta_a} \,.
\end{align}
The value $\theta_{E} = 76^\circ$ can be explained by the spin-orbit interaction with $\lambda/\Delta_a \approx 0.18$, 
the shear strain with $\Xi \sqrt{u_{xz}^2+u_{yz}^2}/\Delta_a \approx 0.26$, or by the combination of both effects. 

With increasing temperature, both the $^4A_2$ and $^4E$ multiplets get thermally populated leading to a reduction of the PL polarization. This depolarization is observed by the comparison between Fig.~\ref{fig2}(d) and Figs.~\ref{fig3}(c,d). At room temperature of Fig.~\ref{fig2}(d), the ratio $\cos 2\theta$ is reduced from 0.96 to 0.72  for  the V2 centers and from $-0.89$ to $-0.26$  for the  V3 centers. This allows us to estimate the energy separation between the multiplets $\Delta_a \approx 60$ meV and $\Delta_a \approx -20$ meV for the V2 and V3 centers, respectively (Supplementary Information).

\section{Temperature dependence of spin properties}

\subsection{Excited-state level anticrossing}

\begin{figure}[t]
\includegraphics[width=.47\textwidth]{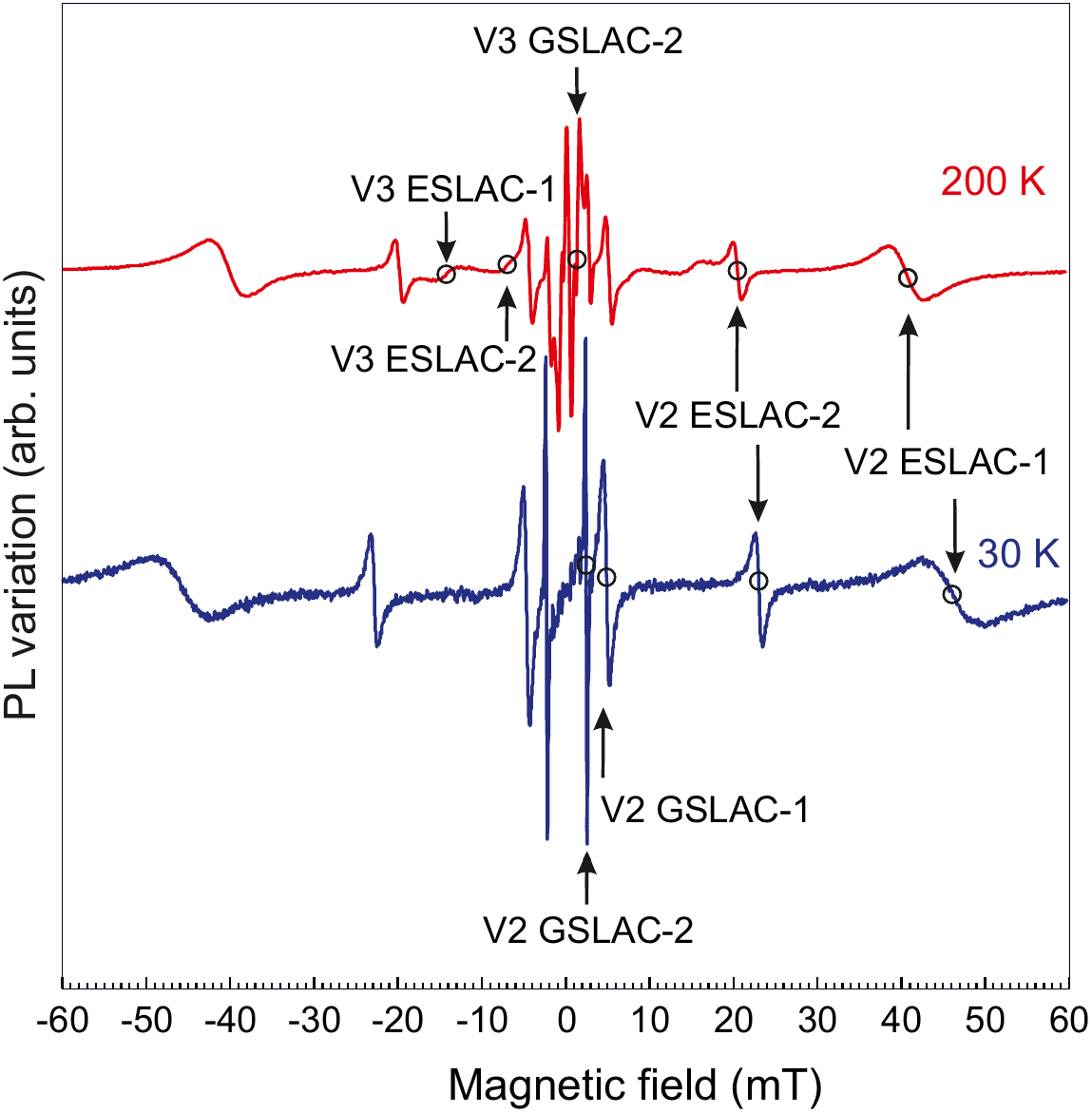}
\caption{PL intensity variation with magnetic field. The arrows indicate the GSLAC and ESLAC for the V3 (left part) and V2 (right part). } \label{fig5}
\end{figure}

To experimentally probe the fine structure of the spin multiplets, we apply external magnetic fields $B$ along the $c$-axis and study how it affects the PL signal. At certain values of  $B$, determined by the zero-field splitting, level anticrossing (LAC) between a pair of spin states occurs \cite{Simin:2016cp}. At these four magnetic fields, corresponding to two ES level anticrossing (ESLAC) and two GS level anticrossing (GSLAC), the spin states are mixed resulting in a resonant change of the PL intensity
 \cite{Anisimov:2016er, Anisimov:2018cp}. The shape of the PL intensity in the vicinity of LAC is determined by the parity and spin dynamics of the corresponding spin levels \cite{Tarasenko:2017ky, PhysRevB.103.014403}. 
Figure~\ref{fig5} shows the magnetic field dependence of the PL for two temperatures.
The narrow GSLAC resonances at low magnetic fields allow us to determine the GS zero-field splitting $2D$ \cite{Simin:2016cp}. The broader ESLAC resonances show clear temperature dependence and we determine the ES zero-field splitting $2D'$ together with the thermal shift $\beta = 2 dD' / dT$ \cite{Anisimov:2016er}. All these parameters are summarized in Table~\ref{LatticePar}. 

 \begin{table}
\caption{\label{LatticePar} Temperature dependence of the zero-field splitting in the GS ($2D$) and ES ($2D'$). The temperature shift is obtained as $\beta = 2 \, dD'/dT$ at room temperature from the ESLAC \cite{Anisimov:2016er}.  }
\begin{ruledtabular}
\begin{tabular}{lcccc}
Center & $2D$ & $2D'$ @ 300 K & $\beta$ @ 300 K \\
\hline
V2 & 128 MHz & 1050 MHz & $-$1.16 MHz/K \\
V3 & $-28$ MHz & $-350$ MHz& 0.84 MHz/K \\
\end{tabular}
\end{ruledtabular}
\end{table}

Surprisingly, the V3 ESLACs in Fig.~\ref{fig5} have small amplitude but are clearly observed even though  the $^4A_2$ quadruplet lies energetically higher than the $^4E$ octuplet due to the V3 inverted ES structure (Fig.~\ref{fig1}). A possible reason is that for non-resonant excitation the V3 centers  reside in the $^4A_2$ state for some time during the excitation-relaxation cycle \cite{Dong:2018uf}. Note that the ESLACs in the $^4E$ octuplet are expected at much higher magnetic fields than those used in our experiments. Furthermore, we find that the V3  GSLAC and ESLAC resonances disappear at lower temperatures (Fig.~\ref{fig5}), indicating that the optical spin pumping mechanism becomes inefficient at certain temperatures.

\subsection{Temperature inversion of the ODMR signal}

 \begin{figure}[t]
\includegraphics[width=.47\textwidth]{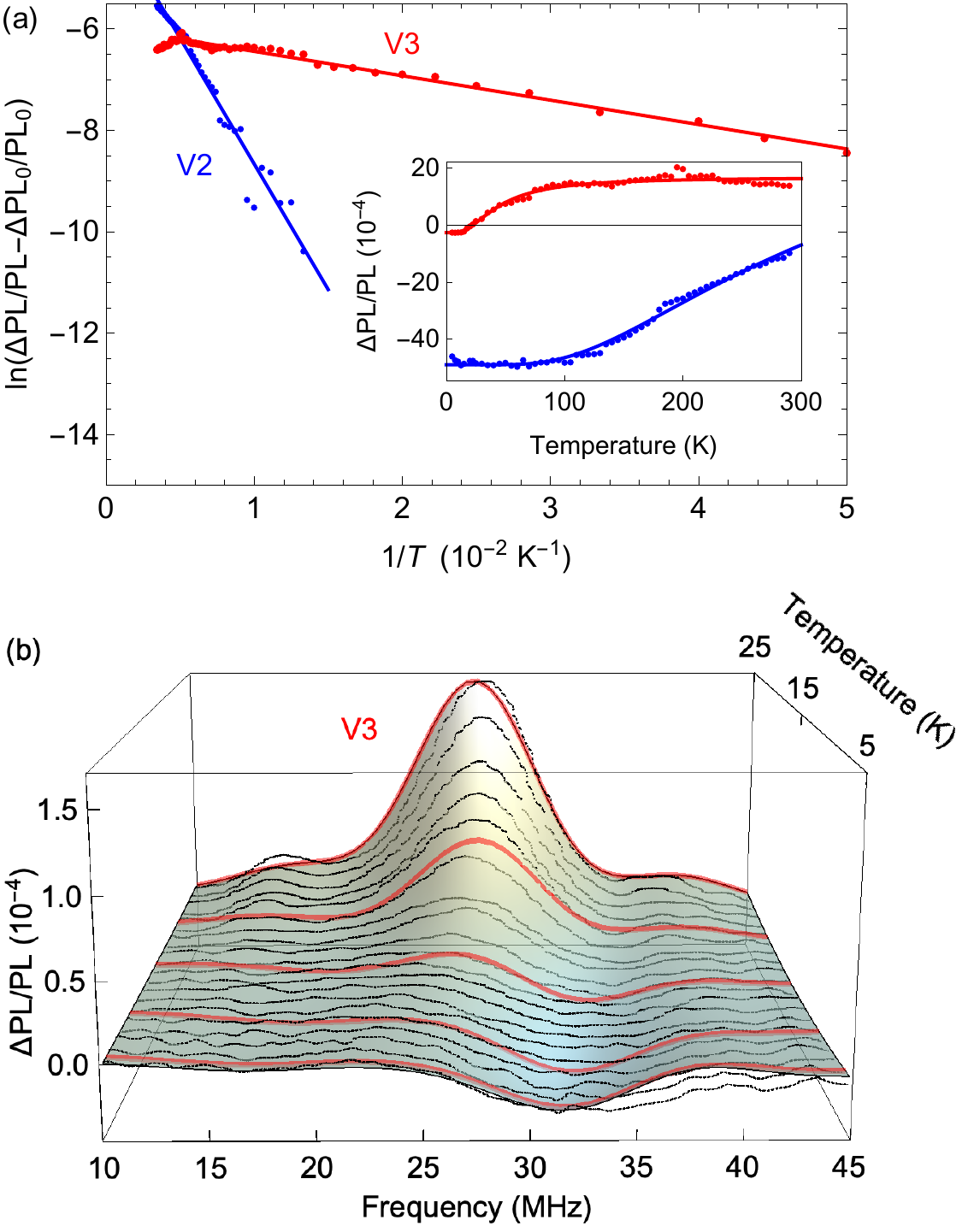}
\caption{(a) Arrhenius plot of the ODMR signal for V2 and V3 centers. The plot was obtained from the ODMR temperature dependencies (inset) after subtracting the constant background $\Delta\text{PL}_0/\text{PL}_0$ corresponding to ODMR at $T=0$. 
(b) Temperature variation of the ODMR signal for the V3 center in the vicinity of the critical temperature, where the sign changes. Dots are the experimental data, the surface and the red lines are the result of the fit after Eq.~\eqref{eq:ODMRT}. The best fit is obtained with the critical temperature $T_c = 16$\,K and $\gamma = 1.4$\,K/MHz.}
\label{fig6}
\end{figure}

To understand the effect of temperature on the spin properties, we measure ODMR spectra in the range of 5 - 300\,K (Fig.~\ref{fig6}). The Arrhenius plot of Fig.~\ref{fig6}(a)  reveals an activation-law behavior for both, the V2 and V3 centers with activation energies of $43\,$meV and $4.1$\,meV, respectively. These values are in agreement, within an order of magnitude, with the values obtained from the polarization dependencies and might be associated with the thermal activation of the either $^4E$ or $^4A_2$ ES in case of the V2 or V3 center, respectively. 

As shown in the inset of Fig.~\ref{fig6}(a), the ODMR contrast for the V2 center has a large negative offset $\Delta\text{PL}_0/\text{PL}_0 = - 5 \times 10^{-3} $  at $T=0$ and remains negative at all temperatures. Interestingly, the offset for  the V3 center $\Delta\text{PL}_0/\text{PL}_0 = - 5 \times 10^{-4} $ at $T=0$ is much smaller. As a result, the ODMR signal for the V3 center changes sign at a critical temperature $T_c$. 

Figure~\ref{fig6}(a) show the ODMR spectra in the vicinity of $T_c$ for the frequency range of the V3 GS resonance. As the temperature drops below 20\,K, the ODMR signal decreases and  the resonance line shape becomes strongly asymmetric. Below 10\,K, the resonance assumes again a symmetric shape but its sign is now negative. The PL intensity (not shown) remains almost constant at the same time. We also study the variation of the electronic paramagnetic resonance (EPR) of the V3 center with temperature, which follows exactly the same behavior (Supplementary Information). 

To describe the temperature variation of the ODMR line shape in the vicinity of $T_c$, we assume  that the optically induced spin quadrupole polarization of the V3 center $d_0$ \cite{Tarasenko:2017ky} is determined by two competing mechanisms, which compensate each other at the critical temperature $T_c$. In the vicinity of this critical temperature, the spin polarization increases linearly as
\begin{align}
d_0(T) \propto T-T_c \,.
\end{align}
In an inhomogeneously broadened ensemble with a variation of the zero-field splitting $2D$, the critical temperature $T_c$ also varies. We suppose that since both variations are likely to be caused by local deformations \cite{doi:10.1063/5.0040936}, so there exists a correlation
\begin{align}
T_c -  T_c^{(0)} =  \gamma(D - D^{(0)}  )
\end{align}
where $T_c^{(0)}$ and $D^{(0)}$ are the average values of the zero-field splitting constant and the critical temperature in the ensemble, respectively, and $\gamma$ is a constant. Then, the ODMR signal becomes
\begin{align} \label{eq:ODMRT}
&\Delta {\rm PL} (\nu) \\\nonumber &\propto \left[T - T_c^{(0)} - \gamma \left(\frac{h\nu}2- D^{(0)}\right)\right] {\rm exp}\left[- \frac{(h\nu/2 - D^{(0)})^2}{2 (\delta D)^2} \right]
\end{align}
where $(\delta D)^2$ is the variance of the zero-field splitting constant. We use Eq.~\eqref{eq:ODMRT} to model the experimental ODMR spectra. From the best fits in Fig.~\ref{fig6}(a), we determine the critical temperature  $T_c = 16$\,K.

\subsection{Rabi oscillations}

\begin{figure}[t]
\includegraphics[width=.47\textwidth]{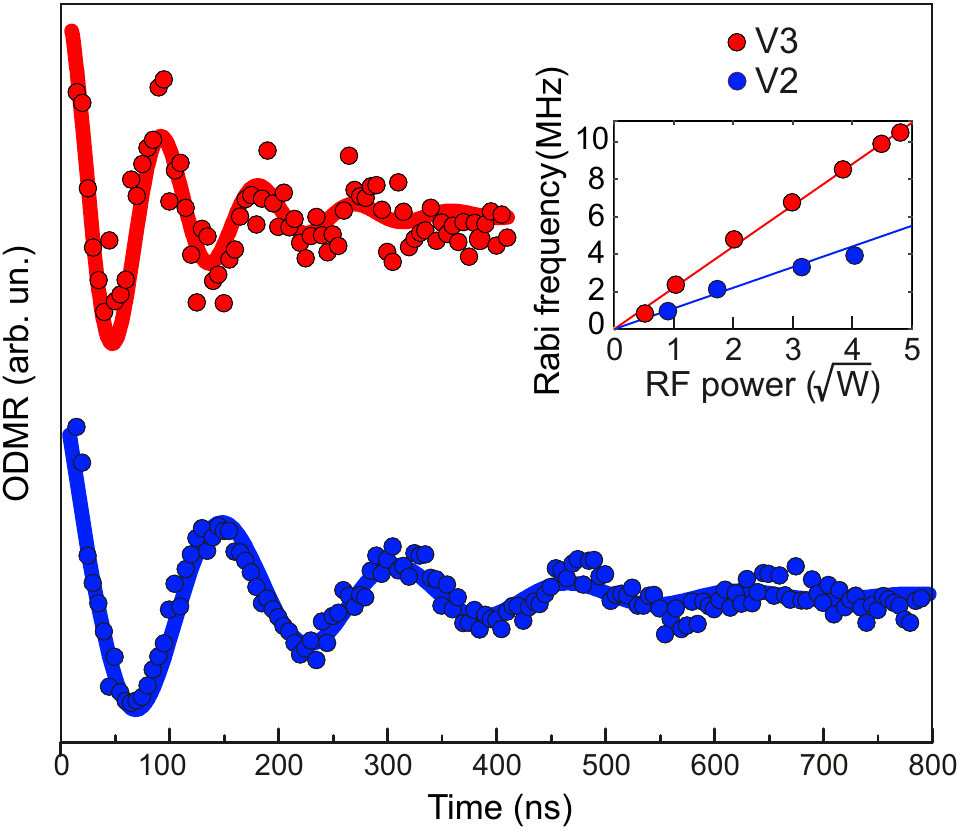}
\caption{Rabi oscillations recorded at 5K for the V2  and V3 centers. Symbols are experimental data and solid lines are fits to Eq.~(\ref{rabi_fit}). Inset: Rabi oscillation frequency as a function of the MW power.}  \label{fig7}
\end{figure}

For quantum spin-photonic applications, it is important to demonstrate the coherent control of the V3 center at low temperature below $T_c$. Figure~\ref{fig7}  shows Rabi oscillations of the V3 spin, which is recorded at the optimal MW frequency of 32~MHz at  $T = 5 \, \mathrm{K}$ according to  Fig.~\ref{fig6}(a) along with a power of 26~W. For comparison, we also measure Rabi oscillations for the V2 spin with the conventional ES structure at the same temperature. In this case, we use a MW frequency of 128 MHz with a power of 20 W, which equals the V2 spin resonance. The experimental data are well fitted to a function
\begin{align}
S_{\mathrm{MW}}(\tau)-S_0(\tau) = A + B \cos (\omega \tau +  \phi) \, e^{-\tau/T_2^*} \,,
\label{rabi_fit}
\end{align}
where $S_{\mathrm{MW}}$ and $S_0$ are the averaged PL signals measured with and without MW pulse, respectively. The obtained inhomogeneous dephasing times $T_2^*$ are $219 \pm16$~ns  $129 \pm 20$~ns for the V2 and V3 ceneters, respectively. These times are similar to earlier reported values for room-temperature experiments in 6H-SiC  \cite{Singh2020,Singh2021,Soltamov2021}. 

\section{Discussion}

The exact values of the spin-orbit interaction and the local deformation are not determined in our experiments. They should be performed on single V3 centers using polarization-resolved PL excitation spectroscopy \cite{Morioka:2020iv}, which is beyond the scope of this work.  Here, we discuss the spin-photon interface for two limits, $\lambda \gg \Xi u_{xz}$ and $\lambda \ll \Xi u_{xz}$. 

The optical selection rules for the dominant spin-orbit interaction are presented in Fig.~\ref{fig4}(d). Particularly, the optical transitions from the lowest, two-fold degenerate ES have the same energy and are circularly polarized $\sigma^{\pm}$ depending on the final GS $| \mp 3/2 \rangle$.  This II-system has some similarities to the trion spin states  \cite{PhysRevB.71.201312} and optical selection rules in semiconductor quantum dots, suggested for the implementation of quantum repeaters  \cite{DeGreve2016}. One of the robust quantum repeater schemes makes use of the entanglement between the spin state and photon polarization. The entanglement protocol consists of several well-documented steps. First, the system is initialized into the $| + 3/2 \rangle$ GS state using resonant GS-ES optical excitation together with a resonant $(- 3/2 \rightarrow -1/2)$ MW field \cite{Nagy:2019fw}. Then, the superposition $| + 3/2 \rangle + | - 3/2 \rangle$ is created by applying a sequence of $\pi/2$ and $\pi$ MW pulses to the dipole-allowed transitions in the GS \cite{Soltamov:2019hr}. Finally, a resonant excitation into the ES with the optical $\pi$ pulse followed by the spontaneous emission generates the spin-photon entangled state $| \sigma^{-} \rangle | + 3/2 \rangle + | \sigma^{+} \rangle | - 3/2 \rangle$. 

In case of the dominant deformation coupling shown in  Fig.~\ref{fig4}(e), the lowest ES state is four-fold degenerate. The optical transitions from this state to the $| \pm 1/2 \rangle$ and $| \pm 3/2 \rangle$ GS are linearly polarized, e.g., $e_x$, and differ in their emission wavelength. This case can also be extended to the spontaneous deformation in the ES due to the Jahn-Teller effect with random linear polarization $e_{x,y}$ of the emission. Such a $\Lambda$-system bears some analogies to the silicon-vacancy center in diamond. It enables memory-enhanced quantum communication and realization of heralded spin-photon gates \cite{Bhaskar:2020gh}. 

The practical implementation of spin-photon repeaters requires indistinguishability of the emitted photons and their high emission rate. The former can be realized with the Stark tuning \cite{Ruhl:2019hs, Lukin:2020jb}. The demonstrated high spectral stability of the V1/V2 centers in SiC in the earlier works \cite{Banks:2019je, Morioka:2020iv} is caused by the identical $^4A_2$ symmetry in the GS and lowest ES \cite{Udvarhelyi:2019eh}, which is not the case for the inverted ES structure. Therefore, the spectral stability of single V3 centers should be benchmarked in future experiments. An improvement of almost two orders of magnitude of the V1/V2 ZPL has been demonstrated by coupling them into photonic structures\cite{Radulaski:2017ic, Bracher:2017il, Lukin:2019fh}. In these experiments, photon collection is performed along the $c$-axis, which is unfavorable for the V1 and V2 centers. Given the  directional emission of the V3 center along the $c$-axis, we expect even a higher photon extraction rate.   

In summary, we report the inverted structure of the excited states of the V3 $\mathrm{V_{Si}}$-center in 6H-SiC, leading to the unusual temperature behaviour of the optical spin pumping. By combining an experimental study with a theoretical model, we establish the selection rules for all optical transitions. It results in the multi-polarized directional emission with favorable orientation for the photon extraction from photonic structures. Furthermore, the inverted ES structure enables robust spin-photon entanglement protocols, which are essential for quantum networks.  Our findings demonstrate that the V3 $\mathrm{V_{Si}}$ in 6H-SiC is a promising spin center for quantum applications, which should stimulate further development of other non-conventional SiC polytypes as a material platform for wafer-scale quantum technology.

\section{Methods}
The 6H-SiC crystals are grown by physical vapor deposition with on-axis orientation. The micropore density is about $5 \, \mathrm{cm^{-2}}$ and the residual dopant concentration below $1 \times 10^{17} \, \mathrm{cm^{-3}}$. The original crystal (3~inches diameter and 100~mm in thik) is diced in rectangular parallelepipeds with a thickness along the $c$-axis of $1.15 \, \mathrm{mm}$ and a base of $2.3 \times 2.3 \, \mathrm{mm^{2}}$. After dicing, all six surfaces of the sample are grounded and polished using a diamond slurry. To create $\mathrm{V_{Si}}$, electron irradiation with an energy of 2~MeV to a fluence of $1 \times 10^{18}  \, \mathrm{cm^{-2}}$ is performed throughout the entire volume of the samples. 

A scheme of our cw and time-resolved ODMR setup is shown in the Supplementary Information. To measure Rabi oscillations, the laser pulse with a duration of 20~ms and a power of 38~mW is used to initialize the $\mathrm{V_{Si}}$ spins. It is followed by a MW pulse of variable duration applied to the sample through Helmholtz coils, producing AC magnetic fields perpendicular to the $c$-axis of the crystal. The PL signal is recorded by the second laser pulse with a duration  $16 \,  \mathrm{\mu s}$. For reference, we repeat the same sequence except for the RF pulse.  The reference is subtracted from the signal to eliminate the background.

\section*{Acknowledgments}
The work was supported by the Russian Foundation for Basic Research, grant Nr. 19-52-12058 as well as by the Deutsche Forschungsgemeinschaft (DFG) grant No.~AS~310/5-1 and in the framework of ICRC, project TRR 160 (Project No.~C7). A.V.P. acknowledges the support from the Russian President Grant No. MK-4191.2021.1.2 and the Foundation "BASIS". Z.S. thanks support from the China Scholarship Council (CSC File No.: 201706220060). 
Theoretical work at the Ioffe Institute was supported by the Russian Science Foundation (grant 19-12-00051).  We thank Manfred Helm for encouragement.

\bibliography{6H-SiC}

%***********************************

\end{document}

\renewcommand{\thefigure}{S\arabic{figure}}
\renewcommand{\theequation}{S\arabic{equation}}

\bibliographystyle{apsrev}
%\bibliographystyle{naturemag}

%\preprint{}

\title{Supplemental Material for:  \\
Inverted fine structure of a 6H-SiC qubit enabling robust spin-photon interface}

\author{I.~D.~Breev$^{1,5}$}
\author{Z.~Shang$^{2,3,5}$}
\author{A.~V.~Poshakinskiy$^{1}$}
\author{H.~Singh$^{4}$}
\author{Y.~Berenc{\'e}n$^{2}$}
\author{M.~Hollenbach$^{2,3}$}
\author{S.~S.~Nagalyuk$^{1}$}
\author{E.~N.~Mokhov$^{1}$}
\author{R.~A. Babunts$^{1}$}
\author{P.~G.~Baranov$^{1}$}
\author{D.~Suter$^{4}$}
\author{M.~Helm$^{2,3}$}
\author{S.~A.~Tarasenko$^{1}$}
\author{G.~V.~Astakhov$^{2}$}
\email[E-mail:~]{g.astakhov@hzdr.de} 
\author{A.~N.~Anisimov$^{1}$}
\email[E-mail:~]{aan0100@gmail.com}

\affiliation{$^1$Ioffe Institute, Polytechnicheskaya 26, 194021 St.~Petersburg, Russia  \\ 
$^2$Helmholtz-Zentrum Dresden-Rossendorf, Institute of Ion Beam Physics and Materials Research, Bautzner Landstr. 400, 01328 Dresden, Germany  \\
$^3$Technische Universit\"{a}t Dresden, 01062 Dresden, Germany \\
$^4$Fakult\"at Physik, Technische Universit\"at Dortmund, 44221 Dortmund, Germany \\ 
$^5$These authors contributed equally to this work}

\maketitle

\tableofcontents

%---------------------------------------------------------------

\section{Methods}
\subsection{ODMR setup}

\begin{figure}[h!]
\includegraphics[width=.7\textwidth]{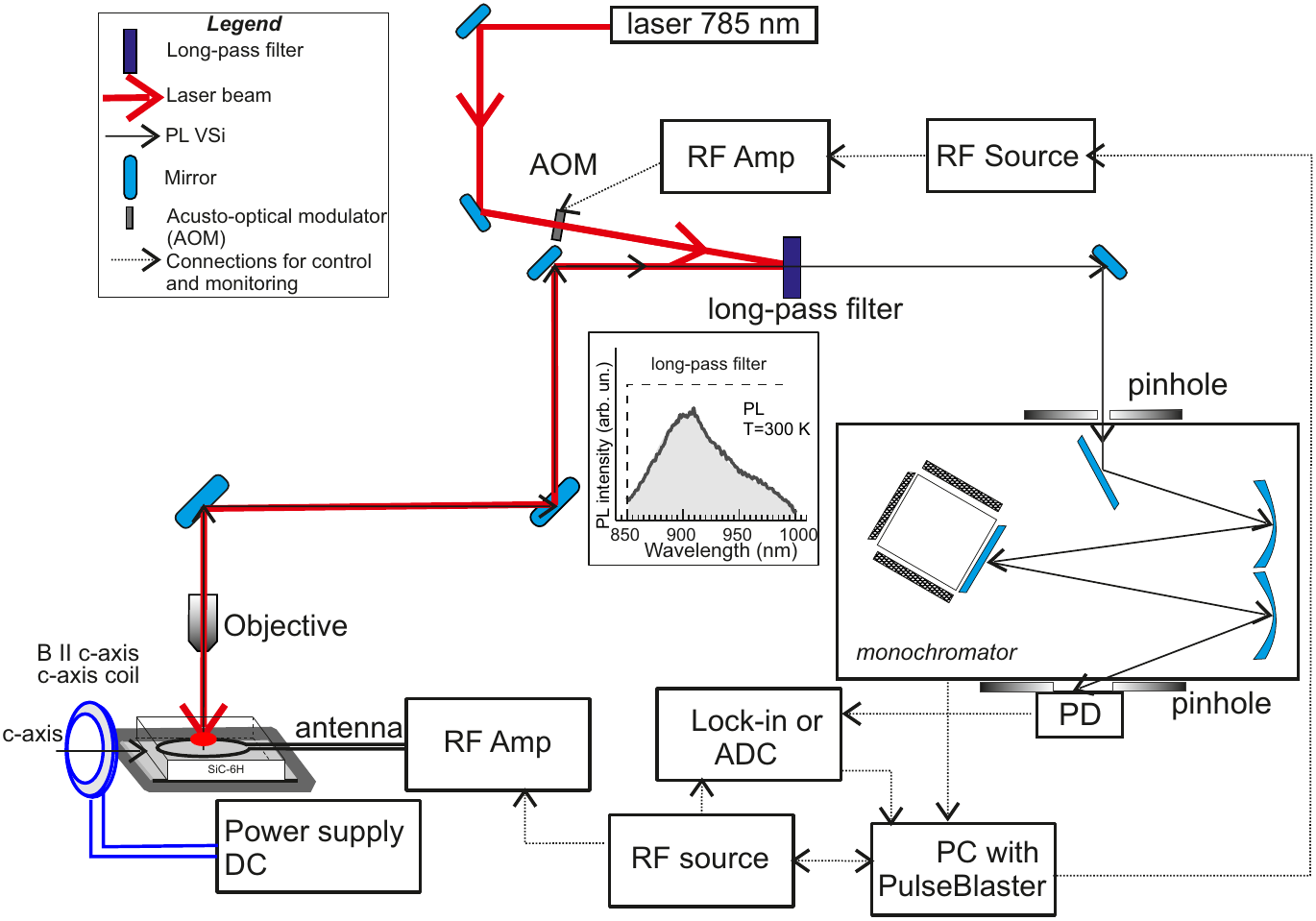}
\caption{(a) Schematic picture of the 6H-SiC sample geometry showing the
crystallographic $c$-axis, crystal sides $n$ and $m$, the PL scanning axes ($a$ and $b$) together with the directions of the PL collection and excitation. (b) PL intensity
dependence on the focal point position along the $a$ axis. (c) PL spectrum of the 6H-SiC sample under 532~nm laser
excitation at room temperature. (d) PL intensity dependence on the focal
point position along the $b$ axis.} \label{figS1}
\end{figure}

Setup used for the cw- and time-resolved ODMR measurements show in Fig.~\ref{figS1}. For cooling the sample, we used a liquid helium flow cryostat (MicrostatHe-R from Oxford instruments). Used a turbopump for creating a vacuum in the cryostat pressure less than $10^{-6}$ mbar (Pfeiffer). Used a 785 nm diode laser as our light source (Thorlabs LD785-SE400), a laser diode controller (LDC202C series) and a temperature controller (TED 200C)). An acousto-optical modulator (NEC model OD8813A) was used for creating the laser light pulses. For applying the static magnetic field to the sample, we used three orthogonal coil-pairs. The PL signal was recorded with an avalanche photodiode (APD) module (C12703 series from Hamamatsu). The signal from the APD was recorded with the USB oscilloscope card (PicoScope 2000 series) during pulse mode ODMR experiments. For cw-ODMR, the signal from APD was recorded with the lock-in (SR830 DSP). For cw-ODMR experiments as an RF source, we used a direct digital synthesizer (DDS AD9915 from Analog Devices). For pulsed ODMR experiments, we used an arbitrary wave generator (AWG DAx14000 from Hunter Micro). An RF signal from the source was amplified using an RF amplifier (LZY-22+ from mini circuits). Was used a digital word generator to generate TTL-pulses (PulseBlaster ESR-PRO PCI card).The RF to the sample was applied in the CW experiments using a 50 $\mu$m wire terminated by a 50-Ohm resistor placed over the sample. For the pulsed mode ODMR experiments, an RF pulse was applied to the sample through a handmade Helmholtz-pair of RF coils with 2.5 mm diameter and 6 turns in each coil from 100 $\mu$m diameter wire terminated with a 50-Ohm resistor. 

\subsection{Angular-dependence measurements}

\begin{figure}[h!]
\includegraphics[width=.89\textwidth]{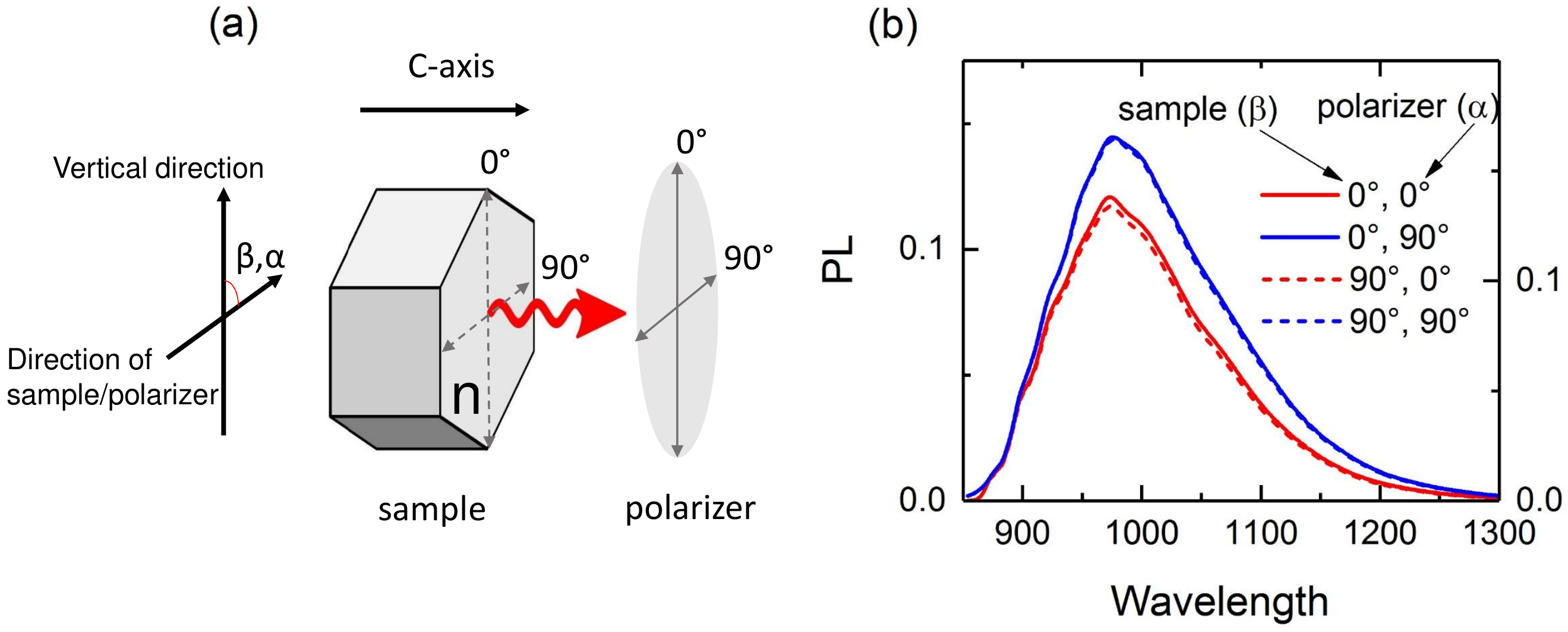}
\caption{(a) Calibration of the angular-dependent detection sensitivity: $\alpha$ is the angle between the polarizer direction and the vertical direction (monochromator slit), $\beta$ is the angle between the sample direction and vertical direction. (b) PL measurements for different $\alpha$ and $\beta$. } \label{figS2}
\end{figure}

When the PL signal passes through optical components (such as monochromator), there is a polarization-dependent intensity loss. Therefore, we perform the calibration of the the angular-dependent detection sensitivity as shown schematically in Fig.~\ref{figS2}(a). The PL intensity $I_{meas}$ of the unpolarized signal $I_{\mathrm{PL}}$ depends on the angle  $\alpha$ between the axis of the linear polarizator and the certain direction (for instance, defined by the monochromator slit) as
\begin{equation}
I_{meas} =  I_{\mathrm{PL}} (1 + A - A \cos 2  \alpha) \,.
 \label{EqS1}
\end{equation}
Here, $1 + 2A$ is the correction factor due to the different sensitivity for $\alpha = 0^\circ$ and $\alpha = 90^\circ$. To take into account the possible intrinsic PL polarization, we perform measurements for two sample orientations with $\beta = 0^\circ$ and $\beta = 90^\circ$ as presented in Fig.~\ref{figS2}(b). The correction factor for different wavelengths ($\lambda$) is obtained as 
\begin{equation}
1 + 2A(\lambda)  =  \frac{I_{0^\circ,90^\circ}(\lambda) + I_{90^\circ,90^\circ}(\lambda)}{I_{0^\circ,0^\circ}(\lambda) + I_{90^\circ,0^\circ}(\lambda)}\,, 
 \label{EqS2}
\end{equation}
where $I_{\beta, \alpha} (\lambda)$ is the PL intensity for the given orientations of the sample ($\beta$) and polarizer ($\alpha$) at a wavelngth $\lambda$. To obtain the angular dependences of the PL signal $I_{\mathrm{PL}} (\alpha)$, the measured intensities $I_{meas} (\alpha)$ are corrected for each $\lambda$ using Eq.~(\ref{EqS1}). 

\begin{figure}[h!]
\includegraphics[width=.89\textwidth]{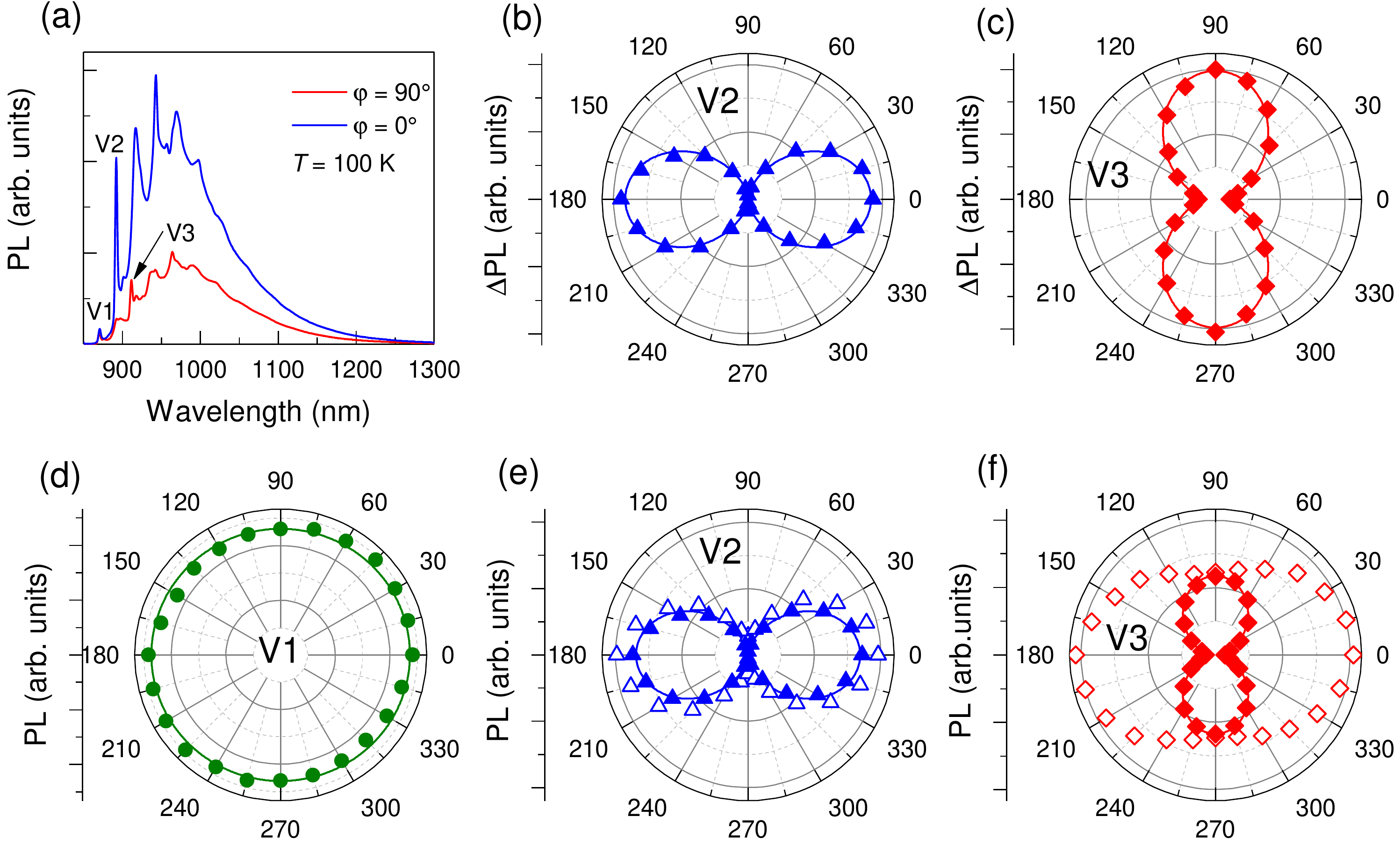}
\caption{(a) PL spectra from the $m$-face at $\varphi_m = 0^\circ$ and $90^\circ$ at 100~K. Here, $\varphi_m$  is the angle between the $c$-axis and the polarizer orientation. (b) and (c) V2 and V3 polar plots obtained from the MW-assisted spectroscopy. (c), (d) and (e) V1, V2 and V3  polar plots obtained from the PL measurements. The open symbols represent the PL intensity at the wavelength of the corresponding ZPL. The solid symbols are the values after the PSB subtraction. The solid lines are fits as described in the text.} \label{figS3}
\end{figure}

Figure~\ref{figS3} presents polar plots of the $\mathrm{V_{Si}}$ in 6H-SiC obtained by two alternative approaches. In the first approach, we use the MW-assisted spectroscopy, which allows to separate overlapping contributions from different color centers \cite{Shang:2021bq}. The angular-dependence measurements are presented in Figs.~\ref{figS3}(b) and (c) at a temperature $T = 100$~K. There are clear angular dependencies of the V2 and V3 $\mathrm{V_{Si}}$ centers, with orthogonally oriented polar plots. In the second approach, we measure the ZPL intensity and subtract the PSB contribution from other centers as presented in Figs.~\ref{figS3}(d)-(f) for the V1, V2 and V3 ZPLs, respectively. Both approaches give the same results. 

\begin{figure}[h!]
\includegraphics[width=.89\textwidth]{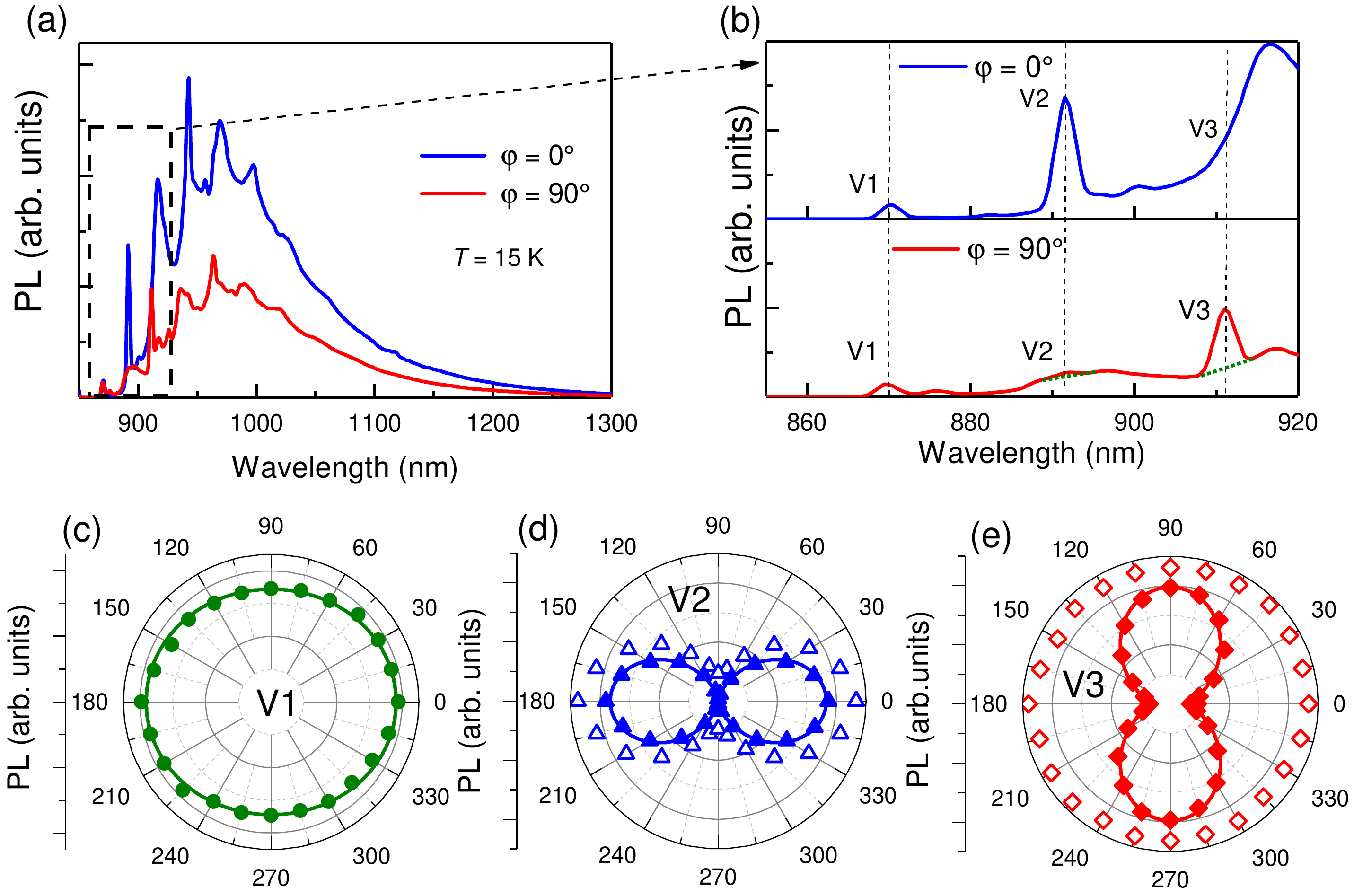}
\caption{PL spectra from the $m$-face at $\varphi_m = 0^\circ$ and $90^\circ$ at 15~K. (b) Zoom into the V1, V2 and V3 ZPLs. (c), (d) and (e) V1, V2 and V3  polar plots obtained from the PL measurements. The open symbols represent the PL intensity at the wavelength of the corresponding ZPL. The solid symbols are the values after the PSB subtraction. The solid lines are fits as described in the text.} \label{figS4}
\end{figure}

Further validation is presented in Fig.~\ref{figS4} for lower temperature $T = 100$~K. In case V1, the $\mathrm{ZPL_{V1}}$ angular dependent  can be directly obtained from the low temperature PL spectra  (Fig.~\ref{figS4}(a)) because there is no contribution from the V2 or V3 ZPLs (Fig.~\ref{figS4}(c)). The PL intensity at the V2 ZPL includes two contributions, the  V2 ZPL and the V1 PSB 
\begin{equation}
\mathrm{ZPL_{V2}} =  \mathrm{PL_{at \, ZPL_{V2}} }  -  \mathrm{PSB_{V1 \, at \, ZPL_{V2}} }  \,. 
 \label{ZPL_V2}
\end{equation}
We assume that the ratio $a = \mathrm{ZPL/PSB}$ is polarization independent and Eq.~(\ref{ZPL_V2}) can be rewritten as 
\begin{equation}
\mathrm{ZPL_{V2}} =  \mathrm{PL_{at \, ZPL_{V2}} }  -  a \mathrm{ZPL_{V1}}  \,. 
 \label{ZPL_aV2}
\end{equation}
The parameter $a$ is obtained from the PL spectrum at $\varphi_m = 90^\circ$, when the contribution from the V2 ZPL is negligible (Fig.~\ref{figS4}(b)). The V2 angular dependence before and after the PSB subtraction is presented in Fig.~\ref{figS4}(d). Similarly, the PL intensity at the V3 ZPL includes three contributions, the V3 ZPL, the V1 PSB and the V2 PSB, which can be rewritten as  
\begin{equation}
\mathrm{ZPL_{V3}} =  \mathrm{PL_{at \, ZPL_{V3}} }  -  b \mathrm{ZPL_{V1}}  - c \mathrm{ZPL_{V2}}\,. 
 \label{ZPL_bcV2}
\end{equation}
The parameters $b$ and $c$ are obtained from the PL spectra of Fig.~\ref{figS4}(b) at $\varphi_m = 90^\circ$ and $\varphi_m = 0^\circ$, respectively. The V3 angular dependence before and after the PSB subtraction is presented in Fig.~\ref{figS4}(e).

%\newpage

\section{Supporting figures}

\begin{figure}[h!]
\includegraphics[width=.5\textwidth]{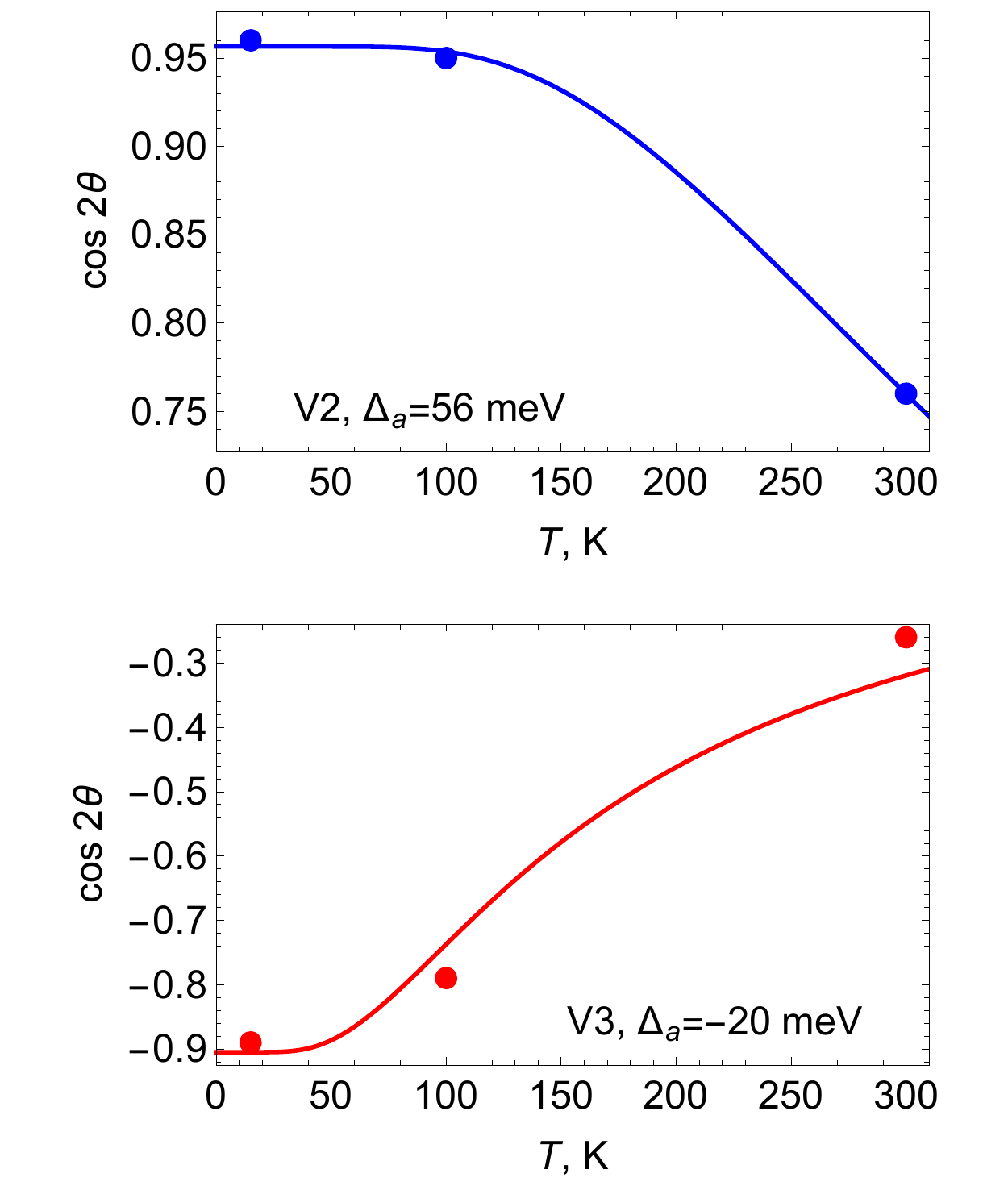}
\caption{
Temperature dependence of the parameter $\cos 2\theta$ extracted from the polarization angular dependence of the V2 and V3 centers PL collected from the the $m$-face. The experimental data (dots) were fitted using the  theoretical equations $\tan^2 \theta (T) = \tan^2\theta(0) + {\rm e}^{-\Delta_a/kT}$ and $\cot^2\theta(T) = \cot^2\theta(0) + {\rm e}^{\Delta_a/kT}$ for V2 and V3 centers, respectively, which yielded the values of 
$\Delta_a$ indicated in the graphs. }
\end{figure}

\begin{figure}[h!]
\includegraphics[width=.89\textwidth]{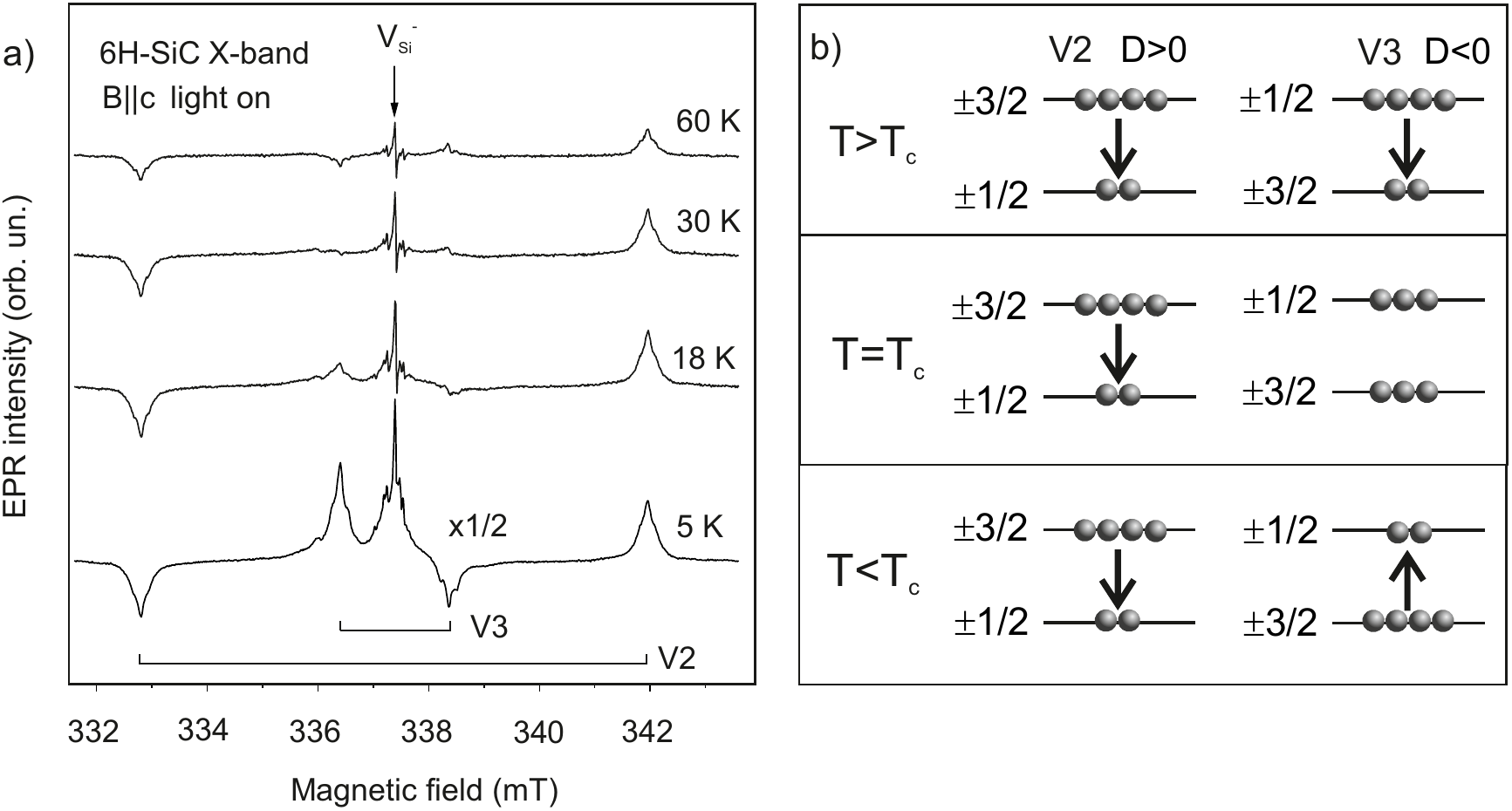}
\caption{(a) Temperature dependence of the X-band EPR spectra of the S = 3/2 family centres in 6H-SiC under optical illumination. Magnetic field is applied along the $c$-axis. Vertical bars indicate the positions of the V2 and V3(V1) EPR lines. (b) Optically induced spin population in the V1, V2 and V3 ground states at different temperatures. } \label{figS5}
\end{figure}

%\section*{Acknowledgments}
%\begin{acknowledgments}
%This work has been supported by...  
%\end{acknowledgments}

%***********************************

\bibliography{6H-SiC_V4}

%***********************************